\documentclass[letterpaper]{article} 
\newcommand{\methodname}{PepTIDE}
\usepackage{aaai2027}  
\usepackage[hyphens]{url}  
\usepackage{graphicx} 
\usepackage{natbib}  
\usepackage{caption} 
\usepackage{algorithm}
\usepackage{algorithmic}
\usepackage{amsmath}
\usepackage{amssymb}
\usepackage{newfloat}
\usepackage{listings}
\DeclareCaptionStyle{ruled}{labelfont=normalfont,labelsep=colon,strut=off} 
\floatstyle{ruled}
\newfloat{listing}{tb}{lst}{}
\floatname{listing}{Listing}

\usepackage{booktabs}

\title{Multi-Scale Temporal Flows for Peptide Trajectory Generation}

\author{
    Xichen Sun\textsuperscript{\rm 1,2},
    Wentao Wei\textsuperscript{\rm 1,3},
    Xiaoxi Zhang\textsuperscript{\rm 1,2}\corresponding,
    Yuedong Yang\textsuperscript{\rm 1,4}\corresponding,
    Jiahua Rao\textsuperscript{\rm 1}\corresponding
}
\affiliations{
    \textsuperscript{\rm 1}School of Computer Science and Engineering, Sun Yat-sen University, Guangzhou, Guangdong, China\\

    \textsuperscript{\rm 2}Shenzhen Loop Area Institute, Shenzhen, Guangdong, China\\ 
   
    \textsuperscript{\rm 3}Pengcheng Laboratory, Shenzhen, Guangdong, China\\

    \textsuperscript{\rm 4}Guangdong Provincial Key Laboratory of Computational Science, Sun Yat-sen University, Guangzhou, Guangdong, China

    \{sunxch7, weiwt8\}@mail2.sysu.edu.cn, \{zhangxx89, yangyd25, raojh7\}@mail.sysu.edu.cn
}
\begin{document}
\nocopyright
\maketitle

\begin{abstract}
Peptides occupy a particularly challenging regime of biomolecular dynamics. Short peptides lack a stable folded core and populate broad conformational ensembles, while cyclization adds ring closure, non-local residue coupling, and stereochemical diversity. Deep generative models have made remarkable progress in emulating molecular-dynamics trajectories directly, yet each is trained on windows cut at a fixed interval and therefore observes the process at a fixed temporal resolution. More fundamentally, none takes that interval as an input, so the physical time a window spans is never represented, leaving the sparse frames on which a peptide crosses between metastable states difficult to learn. Here, we introduce \methodname{}, a multi-scale temporal framework for peptide trajectory generation. The stride of each training window is drawn from a continuous range, so one model observes the same dynamics at many temporal resolutions, and an Adaptive Physical-Time Embedding encodes the relative interval of a window together with the absolute time of each frame, making the shared velocity field interval-aware. All frames are generated jointly through a stochastic-interpolant flow, and pair-aware invariant point attention captures the geometry and topology of both linear and cyclic peptides. On PepMD and a 50-system cyclic-peptide benchmark, \methodname{} reaches state-of-the-art distribution agreement and structural validity. Read in temporal order, its trajectories recover these sparse transition frames and reproduce inter-state fluxes more faithfully, confirming that modeling multiple temporal scales is what makes these rare events accessible.
\end{abstract}


\section{Introduction}
Molecular dynamics (MD) simulation is the workhorse for studying how a biomolecule evolves in physical time at atomic resolution, and its object of interest is the trajectory rather than any single structure. Peptides are a demanding setting for modeling this process. A short linear peptide has no stable tertiary core to fluctuate around, and cyclization removes some degrees of freedom only to add head-to-tail closure, cis--trans peptide-bond isomerization, and stereochemical variation from D-amino acids~\citep{fbm,unisim,cpconfscore,cyclicdiffusion,muco}. Consequently, a single peptide trajectory mixes rotamer flips on picoseconds, backbone dihedral transitions on picoseconds to nanoseconds, and isomerization or whole-chain rearrangement on nanoseconds to microseconds~\citep{tempo,fbm}. 
This flexibility is also what makes peptides valuable: in drug design, where macrocyclic peptides have become a central modality, membrane permeability, target affinity, and the entropic cost of binding are governed by conformational dynamics that no single structure can express~\citep{cpconfscore,cyclicdiffusion}.
 
\begin{figure}[t]
    \centering
    \includegraphics[width=0.85\columnwidth]{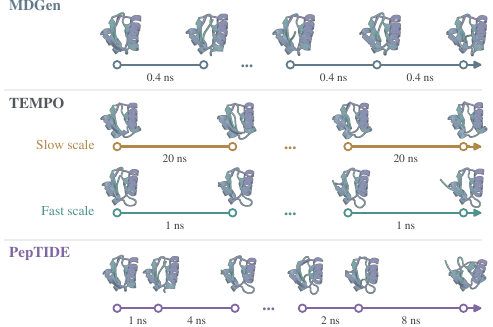}
    \caption{Comparison of temporal sampling strategies.
    MDGen uses a single fixed interval, whereas TEMPO uses two prescribed scales.
    In contrast, \methodname{} is trained across variable physical-time intervals $\Delta \{t_1, t_2,..., t_k \}$.
    }
    \label{fig:temporal_sampling}
\end{figure}

The prohibitive cost of such simulations has motivated deep generative models that emulate them directly, producing whole trajectories rather than isolated conformations without integrating the equations of motion~\citep{mdgen,starmd,confrover,unisim,tempo,ito}. What such a model sees, however, is never the simulation itself but windows of $T$ frames cut from it at a stride fixed during data construction. MDGen learns a stochastic-interpolant flow over an entire trajectory tensor cut at one such stride~\citep{mdgen}; STAR-MD, ConfRover, UniSim, and ATMOS likewise inherit whichever interval their training windows were cut at~\citep{starmd,confrover,unisim,atmos}. A single temporal resolution is an especially poor fit for peptides. Because a peptide fluctuates so freely, what distinguishes one trajectory from another is not its abundant fast motion but the sparse transition frames at which the chain crosses between metastable conformational states. Resolving these frames pulls a model in opposite directions. A short interval captures the fast local motion but leaves the window too brief to reach a basin-to-basin crossing, whereas a long interval spans the slow rearrangement but collapses the intervening frames into a single step~\citep{tempo,ito}. No single stride captures both, and, more fundamentally, none of these models contains a component that takes the interval $\Delta t$ as input, so none can sense the physical timescale or scale the amount of conformational change to it.

To model the timescale at training, several methods condition on it explicitly. TEMPO models slow collective motions and fast local fluctuations in a two-level hierarchy~\citep{tempo}, but this resolves the trajectory at two prescribed scales rather than relating conformational change to $\Delta t$. ITO does make the lag a continuous input, learning self-consistent transition operators at several time resolutions~\citep{ito}, yet only for a single Markovian step. Its lag relates one pair of conformations, so a long trajectory must be assembled by repeated rollout. A generative model is therefore needed that is aware of the physical timescale, learning how much conformational change corresponds to a given time step.


To this end, we introduce \methodname{}, a generative framework for peptide dynamics built on a multi-scale treatment of physical time (Figure~\ref{fig:method_overview}). During training, the stride of each window is drawn from a continuous range instead of being fixed at data construction, so one model observes the same physical process at continually varying temporal resolutions, and every window of $T$ frames is still generated jointly by a stochastic-interpolant flow~\citep{mdgen}. To represent this variation rather than merely be exposed to it, an Adaptive Physical-Time Embedding (APTE) encodes the interval $\Delta t$ of a window as a trajectory-level embedding and the physical time $t_k=k\Delta t$ of each frame as a frame-wise one. The former tells the shared velocity field at which temporal resolution the window was sampled; the latter locates each frame on the simulation-time axis, and both stay distinct from the flow transport time along which samples are drawn. Because the flow emits all $T$ frames at once, the interval reaches every frame within a single computation instead of being consumed one transition at a time. The velocity field is in turn peptide-aware, combining pair-aware invariant point attention over the initial geometry with residue chirality and circular sequence context, so that linear and cyclic peptides are handled by one model.
 
We evaluate \methodname{} on linear-peptide trajectories from PepMD, where it attains the best mean on five of six metrics, and on a 50-system cyclic-peptide benchmark, where it attains the best mean on all of them. On the cyclic benchmark it cuts the six Jensen--Shannon distances---spanning global geometry (\textsc{JS-PwD}, \textsc{JS-Rg}), slow collective coordinates (\textsc{JS-TIC}, \textsc{JS-TIC-2D}), and local torsions (\textsc{JS-Backbone}, \textsc{JS-}$\omega$)---by $28\%$ on average and up to $40\%$ relative to the strongest baseline for each metric, while raising conformational validity \textsc{Val-CA} from $0.620$ to $0.714$ and generating each conformation $24.89\times$ faster than UniSim. That modeling multiple temporal scales is what earns these gains is shown directly. Retraining on a single stride, with the architecture and objective unchanged, raises \textsc{JS-Rg} back to $0.251$ and drops \textsc{Val-CA} to $0.549$~(Table~\ref{tab:interval_coverage_ablation}). Read in temporal order, the generated trajectories further recover the sparse frames on which a peptide crosses between metastable states, rather than oversampling a single basin~(Figure~\ref{fig:rpte_metastable_cases}). Our contributions are summarized as follows:
\begin{itemize}
    \item 
    We propose a multi-scale temporal framework that models peptide trajectories at variable time intervals, drawing the stride of each training window from a continuous range, so the sparse transition frames become learnable alongside the fast local motion.
 
    \item 
    We design an Adaptive Physical-Time Embedding module to equip the trajectory flow with an explicit representation of physical time, encoding the relative interval $\Delta t$ at the trajectory level and the absolute time $t_k=k\Delta t$ at the frame level.
 
    \item 
    \methodname{} achieves state-of-the-art distribution agreement and structural validity on PepMD and on a cyclic-peptide benchmark. Ablations isolate the contributions of APTE and residue-pair geometry, and the generated trajectories reproduce metastable transitions in temporal order.
\end{itemize}

\section{Related Work}
\subsection{From Protein Conformations to Trajectories}
\label{sec:rw_trajectory}
 
Proteins adopt a conformational ensemble rather than a single static structure, and a first line of work learns to sample it, adapting structure predictors through flow matching, learning equilibrium ensembles from large-scale MD, adding physical guidance or structure translation~\citep{alphaflow,bioemu,confdiff,str2str}, or combining likelihood-based proposals with reweighting and Monte Carlo sampling~\citep{boltzmanngenerators,prose}. Such samples, however, carry no time: how long a conformational change takes is outside what they express.
 
A second line therefore models the trajectory itself. Time-coarsened surrogates learn finite-lag transitions and roll them out repeatedly; Timewarp, ITO, and EquiJump study transferable, multi-timescale, and all-atom transitions~\citep{timewarp,ito,equijump}, FBM and UniSim bring learned coarse transitions to peptides and to multiple molecular domains~\citep{fbm,unisim}, and MARS-FM instead models transitions among learned Markov states~\citep{marsfm}. Such rollouts replace many fine MD steps with a coarse learned transition, but condition each step on the current state alone. Multi-frame models close this gap by generating an ordered window jointly: MDGen models the joint distribution of a trajectory with a stochastic-interpolant flow~\citep{mdgen}, TEMPO separates slow collective motions from fast local fluctuations and ConfRover uses causal autoregression to preserve information from previous frames~\citep{tempo,confrover}, while STAR-MD targets stable long-horizon rollout through joint spatiotemporal attention and ATMOS propagates long-range latent dynamics with a state-space model~\citep{starmd,atmos}.
 
Across both lines, however, the interval between consecutive frames is fixed when the training windows are cut, and no component of these models takes it as an input. None can therefore relate the magnitude of conformational change to the time step it spans, which is felt most acutely at the sparse frames on which a molecule crosses between metastable states.
 
\subsection{Peptide and Cyclic-Peptide Dynamics}
\label{sec:rw_peptide}
 
Lacking a stable tertiary core, short peptides interconvert rapidly among broad ensembles, so their behaviour is only meaningful as a time-resolved trajectory rather than as a set of structures. Cyclization compounds this: the head-to-tail bond couples the sequence termini into a global ring-closure constraint, and non-canonical chirality widens the accessible stereochemical space.
 
Work on cyclic peptides has consequently developed along a structural line. 
AfCycDesign and HighFold modify relative positional encodings to represent circular sequence distance, while CP-Composer decomposes cyclic constraints into linear subproblems~\citep{afcycdesign,highfold,cpcomposer}; RINGER and MD-trained cyclic diffusion models generate sequence-conditioned conformational ensembles~\citep{ringer,cyclicdiffusion}. These methods respect the ring but produce structures or unordered ensembles, never trajectories.
 
The trajectory models of the previous section face the converse problem: none of them represents the head-to-tail bond, and on cyclic peptides this is not a minor omission---MDGen and TEMPO generate almost no ring-closed frames at all (Table~\ref{tab:cyclic_main}). Cyclic-peptide dynamics thus falls between the two lines of work, requiring a model that generates ordered trajectories over an explicit physical timescale while respecting ring closure and chirality.

\section{Methodology}
\label{sec:method}



\begin{figure*}[t]
    \centering
    \includegraphics[width=\textwidth]{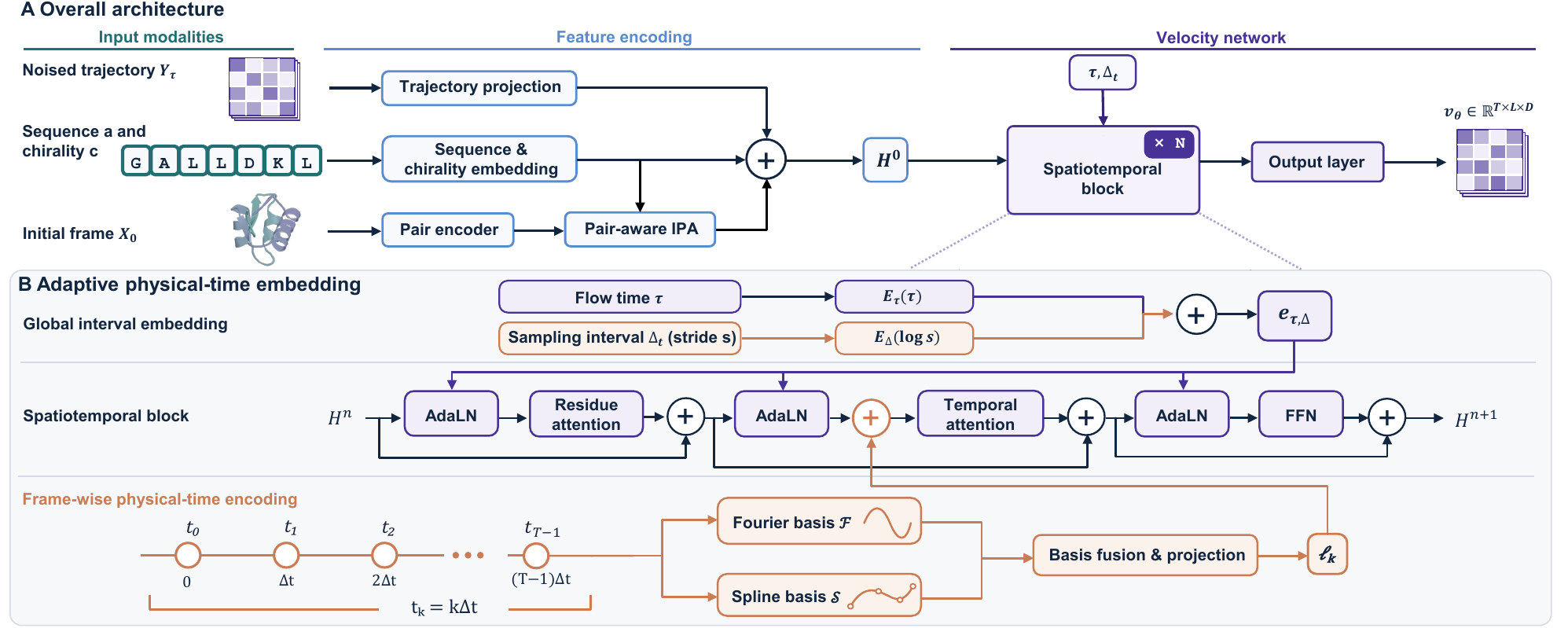}
    \caption{Overview of PepTIDE. \textbf{(A)} The noised trajectory, sequence,
    chirality, and initial geometry are fused into residue--frame states and
    mapped to a whole-trajectory velocity field. \textbf{(B)} The sampling
    interval conditions the spatiotemporal block globally, while the physical-time embeddings of $t_k=k\Delta t$ are injected before temporal attention.}
    \label{fig:method_overview}
\end{figure*}

\subsection{Preliminaries}
\label{sec:method_problem}

\paragraph{Problem formulation.}
Let $\mathbf{a}=(a_1,\ldots,a_L)$ denote a peptide sequence of $L$ residues, and let $\mathbf{c}=(c_1,\ldots,c_L)$ denote residue chirality when available.
A trajectory contains \(T\) conformations \(\mathcal X=(X_0,\ldots,X_{T-1})\) separated by a sampling interval \(\Delta t\), with frame-wise simulation-time coordinates $t_k=k\Delta t\ , k\in\{0,\ldots,T-1\}$.
Given the initial conformation $X_0$, the task is to model
\begin{equation}
    p_{\theta}\!\left(
      X_{1:T-1}\mid X_0,\mathbf{a},\mathbf{c},\Delta t
    \right).
    \label{eq:conditional_task}
\end{equation}
In implementation, the flow operates on a $T$-frame tensor whose $k=0$ slot
is an auxiliary reconstruction of $X_0$. A separate clean encoding of $X_0$
conditions the network, while the remaining slots represent future frames.
Unless otherwise specified, $\Delta t$ and $t_k$ are expressed in base-frame units; their mapping to an absolute physical unit is dataset dependent.

\paragraph{Protein conformation representation.}
Following standard protein representations~\citep{alphafold2,framediff,mdgen,tempo}, we describe residue $l$ at frame $k$ by a backbone rigid transform $g_k^l\in\mathrm{SE}(3)$ and seven backbone and side-chain torsions $\boldsymbol{\varphi}_k^l\in\mathbb{T}^7$. 
To remove dependence on the arbitrary global pose, we anchor motion to the observed initial conformation via $h_k^l=(g_0^l)^{-1}g_k^l$, following the single-key-frame construction of MDGen~\citep{mdgen}. 
Encoding $h_k^l$ by a unit quaternion $q^+(h)$ and a translation vector $\mathbf{t}(h)$, and each torsion on the unit circle as $\mathbf{s}(\varphi)=(\sin\varphi,\cos\varphi)$, gives the residue--frame token
\begin{equation}
    \mathbf{y}_k^l
    =\left[
        q^+(h_k^l),\,
        \mathbf{t}(h_k^l),\,
        \{\mathbf{s}(\varphi_{k,j}^l)\}_{j=1}^{7}
      \right]
    \in\mathbb{R}^{21}.
    \label{eq:trajectory_token}
\end{equation}
Stacking all tokens gives the trajectory tensor $\mathbf{Y}\in\mathbb{R}^{T\times L\times D}$ with $D=21$. This initial-relative representation is invariant to global rigid transformations while retaining the torsions needed to reconstruct all-atom coordinates; canonicalization and decoding details are deferred to Appendix A.1.

\subsection{Rescaling the Physical Timescale}
\label{sec:method_rescale}

Existing joint whole-trajectory generation remains fixed-resolution: in MDGen, the trajectory tensor is cut from each simulation at one fixed stride chosen during data construction, so the physical time spanned by a generated window is a property of the training data rather than a requestable quantity~\citep{mdgen}.
We take the opposite view and treat the sampling interval as a quantity to be sampled during training, exposing a single model to a range of temporal resolutions drawn from the same trajectories.

Given a base trajectory whose consecutive frames are separated by an interval $\Delta t_{\mathrm{base}}$, we draw a feasible integer stride $s$ and subsample the trajectory to form a training window
\begin{equation}
    \mathcal{X}^{(s)}
      =\left(X_0,X_s,X_{2s},\ldots,X_{(T-1)s}\right),
    \quad
    \Delta t=s\,\Delta t_{\mathrm{base}}.
    \label{eq:multi_interval_window}
\end{equation}
The stride $s$ is sampled approximately uniformly in log space over a dataset-specific range, so that resolutions spanning multiplicative scales are seen with comparable frequency.
This construction decouples the tensor length $T$ from the physical span it represents: with $T$ fixed, a small $\Delta t$ yields a short, finely resolved window that captures fast local motion, whereas a large $\Delta t$ yields a long, coarsely resolved window that reaches basin-to-basin rearrangements.

Because a window carries no intrinsic record of its own resolution, the sampled interval $\Delta t$ is retained alongside the trajectory and supplied to the generator as an explicit coordinate, turning the physical timescale into a control variable rather than a fixed property of the data. 
This is what separates \methodname{} from prior work along the temporal axis: MDGen and related whole-trajectory models fix one resolution at data-construction time~\citep{mdgen,atmos}; TEMPO factorizes motion into two discrete timescale bands rather than a continuous axis~\citep{tempo}; and ITO exposes the lag but only within a Markovian transition, so a long trajectory must be assembled by repeated rollout~\citep{ito}. 
In contrast, \methodname{} makes the interval a continuous, requestable coordinate of a single flow that generates the whole trajectory at once.

\subsection{Timescale-Aware Trajectory Flow Matching}
\label{sec:method_flow}
Rather than learning a one-step transition operator, we regard the trajectory tensor $\mathbf{Y}$ as a single random variable and learn a continuous transport from a simple Gaussian noise to its conditional data distribution. 

For each trajectory $\mathbf{Y}$, we sample an equally shaped noise $\boldsymbol{\epsilon}\sim\mathcal{N}(\mathbf{0},\mathbf{I})$ and a scalar flow time $\tau\sim\mathcal{U}(0,1)$, shared across all frames. 
We connect the noise and data samples using the generalized variance-preserving path
\begin{align}
    \mathbf{Y}_{\tau}
        &= \alpha(\tau)\mathbf{Y}
         + \sigma(\tau)\boldsymbol{\epsilon},
         \label{eq:flow_path}
\end{align}
where $\alpha(\tau)=\sin\!\left(\frac{\pi\tau}{2}\right)$ and $\sigma(\tau)=\cos\!\left(\frac{\pi\tau}{2}\right)$. 
Thus, $\mathbf{Y}_0=\boldsymbol{\epsilon}$ and $\mathbf{Y}_1=\mathbf{Y}$.

Differentiating Eq.~\eqref{eq:flow_path} with respect to $\tau$ gives the sample-wise target velocity
\begin{equation}
    \mathbf{U}_{\tau}
      = \frac{\partial\mathbf{Y}_{\tau}}{\partial\tau}
      = \dot{\alpha}(\tau)\mathbf{Y}
        + \dot{\sigma}(\tau)\boldsymbol{\epsilon}.
    \label{eq:velocity_target}
\end{equation}
As this target is available in closed form, it can be evaluated directly for every sampled pair $(\mathbf{Y},\boldsymbol{\epsilon})$ without simulating an intermediate molecular trajectory. 
Let $\mathcal{C}_{\Delta} = \left( \mathbf{X}_0,\mathbf{a},\mathbf{c},\Delta t, \{t_k\}_{k=0}^{T-1} \right)$ denote the conditioning information.
Given the noised trajectory $\mathbf{Y}_{\tau}$, $\tau$, and $\mathcal{C}_{\Delta}$, the conditional velocity network $v_{\theta}$ is trained using
\begin{equation}
    \mathcal{L}_{\mathrm{FM}}
    = \mathbb{E}_{\mathbf{Y},\boldsymbol{\epsilon},\tau}\left[
        \operatorname{MSE}\!\left(
            v_{\theta}(\mathbf{Y}_{\tau},\tau,\mathcal{C}_{\Delta}),
            \mathbf{U}_{\tau}
        \right)
    \right].
    \label{eq:flow_loss}
\end{equation}
The MSE is evaluated only over valid trajectory components, excluding padding and undefined structural channels.


As illustrated in Figure~\ref{fig:method_overview}B, Adaptive Physical-Time Embedding (APTE) conditions the joint trajectory flow on an adaptive physical timescale at two complementary levels. 
The global interval embedding encodes the sampling interval \(\Delta t\) and contributes to the global AdaLN condition throughout the velocity network. 
The frame-wise embeddings of the induced physical-time coordinates $t_k=k\Delta t$ are injected before temporal attention. 
Both levels describe simulation time and remain distinct from the flow-matching time \(\tau\).

\paragraph{Trajectory-level interval embedding.}
Tensors with the same shape may correspond to different sampling resolutions, so the requested interval is supplied directly to the velocity network. We encode the global sampling interval as
\begin{equation}
    \mathbf{e}_{\tau,\Delta}
    =
    E_{\tau}(\tau)
    +
    E_{\Delta}\!\left(
        \log s
    \right),
    \label{eq:global_interval_condition}
\end{equation}
where $E_{\tau}$ and $E_{\Delta}$ are independent scalar-time embedders consisting of sinusoidal features followed by two-layer MLPs. Each velocity block and the output head use $\mathbf{e}_{\tau,\Delta}$ as the conditioning input to their respective adaptive layer normalization (AdaLN) modules.

\paragraph{Frame-Wise Physical-Time Embedding.}
Global conditioning specifies the spacing between frames but does not identify their locations within the trajectory. 
For frame $k$, we therefore encode each relative coordinate $t_k=k\Delta t$ with a continuous time representation~\citep{time2vec,lete}. 
With $u_k= \log(1+t_k)$ in the log-time configuration, the frame embedding is
\begin{equation}
    \boldsymbol{\ell}_k
      =\tanh(g_t)P_t
        \left[
          \mathcal{F}(W_{\mathcal{F}}u_k);\,
          \mathcal{S}(W_{\mathcal{S}}u_k)
        \right],
    \label{eq:continuous_time_encoding}
\end{equation}
where $P_{t}$ is a learnable projection, $g_{t}$ is a scalar gate initialized at zero, and $\mathcal{F}$ and $\mathcal{S}$ denote learnable Fourier and spline basis functions, respectively. The embedding $\boldsymbol{\ell}_k$ is added before temporal attention in every block.
Thus, $\Delta t$ controls the trajectory globally, whereas $t_k$ supplies a frame-specific continuous coordinate; both remain distinct from $\tau$. 
The complete encoder definition is provided in Appendix C.

\subsection{Peptide-Aware Spatiotemporal Velocity Network}
\label{sec:method_network}

As shown in Figure~\ref{fig:method_overview}A, we parameterize the conditional velocity field $v_\theta$ using a pair-aware spatiotemporal Transformer. 
The perturbed trajectory $\mathbf{Y}_\tau$ is projected into residue--frame tokens and augmented with amino-acid, chirality, and clean initial-conformation features. 
We denote the resulting residue--frame hidden tensor by
$\mathbf H^{(0)}\in\mathbb R^{T\times L\times d}$, with the $n$-th
velocity block mapping $\mathbf H^{(n)}$ to $\mathbf H^{(n+1)}$.
To provide structural context, we construct static residue-pair features from the pseudo-$\beta$ geometry of the initial conformation and encode them together with the initial residue frames using an IPA module~\citep{alphafold2}. The resulting representation is shared across all generated frames.
Each velocity block alternates intraframe residue attention with interframe temporal attention. The global condition $\mathbf{e}_{\tau,\Delta}$ modulates the blocks through adaptive layer normalization, while the frame-specific vectors $\{\boldsymbol{\ell}_k\}_{k=0}^{T-1}$ are added before temporal attention. An output head maps the final hidden states to velocities for all trajectory tokens simultaneously. 
Detailed illustrations of each module are provided in Appendix B.

\subsection{Training and Sampling}
\label{sec:method_sampling}

\paragraph{Staged training.}
We optimize the masked velocity-matching objective in Eq.~\eqref{eq:flow_loss} using a staged schedule. 
For cyclic peptides, we first train the model at a fixed sampling stride and then introduce the global interval condition and frame-wise continuous-time encoding during variable-stride training. 
For PepMD, following UniSim~\citep{unisim}, we initialize from an ATLAS-pretrained protein model, adapt it to PepMD at a fixed stride, and subsequently continue training over variable strides with the same temporal conditions. 
This fixed-to-variable schedule preserves the learned initial structure-conditioned dynamics while introducing explicit control over temporal resolution.

\paragraph{Sampling.}

At inference, we sample a \(T\)-frame Gaussian tensor and integrate the learned flow conditioned on \(X_0\), sequence, chirality, and the requested interval \(\Delta t\). 
The clean \(X_0\) also serves as the reference for structural decoding but does not overwrite the generated \(k=0\) slot.
All reported experiments use one jointly generated 1,000-frame window and do not concatenate rollouts. Additional decoding and integration details are provided in Appendices A.1 and F.3.

\section{Experiments}
\label{sec:experiments}
\subsection{Experimental Setup}
\label{sec:exp_setup}



\paragraph{Datasets and baselines.}
We evaluate on two peptide datasets. PepMD provides OpenMM trajectories for 283 peptides of 3--10 residues, and we use the 14-peptide test split of FBM~\citep{fbm,unisim}; CPconf\_score provides $1\,\mu\mathrm{s}$ trajectories for 250 head-to-tail cyclic peptides, split by sequence into 225/25 training/validation systems with an external 50-peptide test set~\citep{cpconfscore}. Each method generates 1,000 frames per test peptide. On PepMD, we follow the protocol of UniSim and compare with FBM, Timewarp, ITO, Score Dynamics, UniSim, and MDGen~\citep{timewarp,ito,scoredynamics,unisim,mdgen}; on the cyclic benchmark we compare with four state-of-the-art generators spanning peptide- and protein-scale dynamics, namely MDGen, TEMPO, UniSim, and MARS-FM~\citep{marsfm}. All methods use the same split and evaluation pipeline; further details are given in Appendix D.

\paragraph{Implementation.}
We measure distributional fidelity using JS distances over pairwise C$\alpha$ distances (\textsc{PwD}), radius of gyration (\textsc{Rg}), and one- and two-dimensional slow TICA coordinates (\textsc{TIC}, \textsc{TIC-2D})~\citep{fbm,unisim}.
\textsc{Val-CA} measures structural validity, and \textsc{Contact} measures residue-contact error on PepMD.
For cyclic peptides, \textsc{JS-Backbone} and \textsc{JS-}$\omega$ additionally assess backbone and peptide-bond torsions~\citep{mdgen,cpconfscore}.
Windows hold 1,000 frames, with strides sampled over $[10,64]$ on PepMD and $[1,16]$ on cyclic peptides, and evaluation at strides 10 and 2.
Appendices E and F give metric definitions, hyperparameters, and hardware.

\begin{table}[t]
    \centering
    \small
    \setlength{\tabcolsep}{4.2pt}
    \resizebox{\columnwidth}{!}{%
    \begin{tabular}{lcccccc}
        \toprule
        Method &
        \multicolumn{4}{c}{JS distance $\downarrow$} &
        \textsc{Val-CA} $\uparrow$ &
        \textsc{Contact} $\downarrow$ \\
        \cmidrule(lr){2-5}
        & \textsc{PwD} & \textsc{Rg} & \textsc{TIC} & \textsc{TIC-2D} & & \\
        \midrule
        FBM
        & 0.361/0.165 & 0.411/0.224 & \underline{0.510}/0.124
        & 0.736/0.065 & 0.539/0.111 & 0.205/0.105 \\
        Timewarp
        & 0.362/0.095 & 0.386/0.120 & 0.514/0.110
        & 0.745/0.061 & 0.028/0.020 & 0.195/0.051 \\
        ITO
        & 0.367/0.077 & 0.371/0.131 & \textbf{0.495}/0.126
        & 0.748/0.055 & 0.160/0.186 & 0.174/0.099 \\
        SD
        & 0.727/0.089 & 0.776/0.087 & 0.541/0.113
        & 0.782/0.042 & 0.268/0.266 & 0.466/0.166 \\
        MDGen
        & 0.344/0.176 & 0.339/0.223 & 0.553/0.135
        & 0.732/0.118 & \underline{0.695}/0.319 & 0.180/0.117 \\
        UniSim
        & \underline{0.328}/0.149 & \underline{0.330}/0.189
        & \underline{0.510}/0.124 & \underline{0.731}/0.074
        & 0.575/0.139 & \underline{0.157}/0.088 \\
        \midrule
        \methodname{}
        & \textbf{0.312}/0.103 & \textbf{0.298}/0.135 & 0.538/0.119
        & \textbf{0.708}/0.115 & \textbf{0.867}/0.228
        & \textbf{0.133}/0.073 \\
        \bottomrule
    \end{tabular}
    }
    \caption{Results on the 14-peptide PepMD test set. Entries are
    mean/std across peptides. The best and second-best values are shown in bold and underlined.}
    \label{tab:pepmd_main}
\end{table}

\subsection{Results on the PepMD Benchmark}
\label{sec:exp_pepmd}

Table~\ref{tab:pepmd_main} reports the 14-peptide PepMD test split, on which \methodname{} attains the best mean on five of the six metrics. The clearest margins are in structural quality: C$\alpha$ validity rises from $0.695$ for the best baseline to $0.867$, and the contact error falls from $0.157$ to $0.133$, a $15\%$ reduction.
These two metrics separate the methods well, and the ordering is stable---the runner-up on \textsc{Val-CA} (MDGen, $0.695\!\pm\!0.319$) is also the least consistent across peptides, whereas \methodname{} combines the highest mean with a smaller spread.
Fewer clashes and chain breaks, together with more accurate residue-contact frequencies, indicate that the generated frames are usable as conformations rather than merely close in aggregate statistics.

The distributional metrics move in the same direction but by smaller margins: against UniSim, the strongest baseline, JS-PwD improves from $0.328$ to $0.312$ and JS-Rg from $0.330$ to $0.298$. With only 14 test peptides of widely varying difficulty, the across-peptide standard deviations ($\pm0.10$--$0.19$) exceed these margins, so we also compare the methods with UniSim peptide by peptide: \methodname{} improves JS-PwD on 11/14 and JS-Rg on 10/14 systems (Wilcoxon signed-rank $p=$0.013 for both metrics), indicating a systematic rather than a system-specific gain.
\methodname{} likewise attains the lowest JS-TIC-2D ($0.708$), while its one-dimensional JS-TIC ($0.538$) does not surpass ITO ($0.495$); the two are not redundant, since \textsc{TIC} averages discrepancies along each slow coordinate separately whereas \textsc{TIC-2D} measures occupancy of their joint support, so a model can cover the joint landscape more completely while remaining less well calibrated along a single axis. Overall, \methodname{} is clearly ahead on structural validity and contact fidelity, and modestly but consistently ahead on global conformational geometry.

\subsection{Results on the Cyclic-Peptide Benchmark}
\label{sec:exp_cyclic}

We use the cyclic-peptide benchmark to test whether a model handles ring closure and non-canonical stereochemistry, and whether it transfers to chemically distinct systems: the 50 test peptides are experimentally characterized, whereas the training pool is randomly generated.

As shown in Table~\ref{tab:cyclic_main}, \methodname{} obtains the best mean in every column among the five evaluated methods, reducing the six JS distances relative to the strongest baseline for each metric by $28\%$ on average. The improvement is broad rather than concentrated in one representation: the largest gains are in global geometry (\textsc{JS-Rg}, $0.343\!\rightarrow\!0.204$) and in the slow collective coordinates (\textsc{JS-TIC}, $0.353\!\rightarrow\!0.232$), both by $34$--$41\%$, while \textsc{Val-CA} rises by $9.4$ percentage points ($0.620\!\rightarrow\!0.714$). Notably, the strongest baseline is not the same in every column---UniSim leads on the slow coordinates and \textsc{Val-CA}, MARS-FM on global geometry and \textsc{JS-Backbone}---so no single prior method is simultaneously competitive across the three families of statistics. The smallest gain is on \textsc{JS-TIC-2D} ($11\%$), the metric with the highest absolute value for every method: joint occupancy of the two slowest coordinates is the most demanding statistic in the set, and its residual error indicates that reproducing the full two-dimensional slow landscape remains open. The near-zero \textsc{Val-CA} of MDGen ($0.000$) and TEMPO ($0.026$) reflects the metric rather than a failed adaptation: neither model represents the head-to-tail bond, so almost every generated frame violates the ring-closure condition. On this benchmark \textsc{Val-CA} therefore indicates whether cyclization is modeled at all, and should be read together with the JS distances.


These results are consistent with the design of \methodname{}, although the main table alone does not attribute them to individual components. The gains on \textsc{JS-TIC} and \textsc{JS-TIC-2D} are consistent with generating the whole window jointly over an interval-indexed timescale, while the improvements on the torsional metrics and \textsc{Val-CA} reflect the strength of the pair-aware velocity field in capturing residue-pair geometry. 
The ablations below provide additional evidence for these temporal and pair-aware design choices.

\begin{table}[t]
    \centering
    \small
    \setlength{\tabcolsep}{5pt}
    \resizebox{\columnwidth}{!}{%
    \begin{tabular}{lccccccc}
        \toprule
        Method & \multicolumn{6}{c}{JS distance $\downarrow$} & \textsc{Val-CA} $\uparrow$ \\
        \cmidrule(lr){2-7}
        & \textsc{PwD} & \textsc{Rg} & \textsc{TIC} & \textsc{TIC-2D} & \textsc{Backbone} & $\omega$ & \\
        \midrule
        MDGen   & 0.635 & 0.815 & 0.614 & 0.788 & 0.632 & 0.604 & 0.000 \\
        TEMPO   & 0.571 & 0.662 & 0.619 & 0.817 & 0.531 & 0.623 & 0.026 \\
        UniSim  & 0.386 & 0.419 & \underline{0.353} & \underline{0.597} & 0.333 & \underline{0.266} & \underline{0.620} \\
        MARS-FM & \underline{0.301} & \underline{0.343} & 0.384 & 0.648 & \underline{0.306} & 0.272 & 0.556 \\
        \methodname{} & \textbf{0.201} & \textbf{0.204} & \textbf{0.232} & \textbf{0.531} & \textbf{0.225} & \textbf{0.208} & \textbf{0.714} \\
        \bottomrule
    \end{tabular}
    }
    \caption{Results on the 50-cyclic-peptide test set. Entries are means across peptides; best and second-best values are shown in bold and underlined.}
    \label{tab:cyclic_main}
\end{table}

\subsection{Ablation Study}
\label{sec:exp_ablation}

\begin{figure}[t]
    \centering
    \includegraphics[width=\columnwidth]{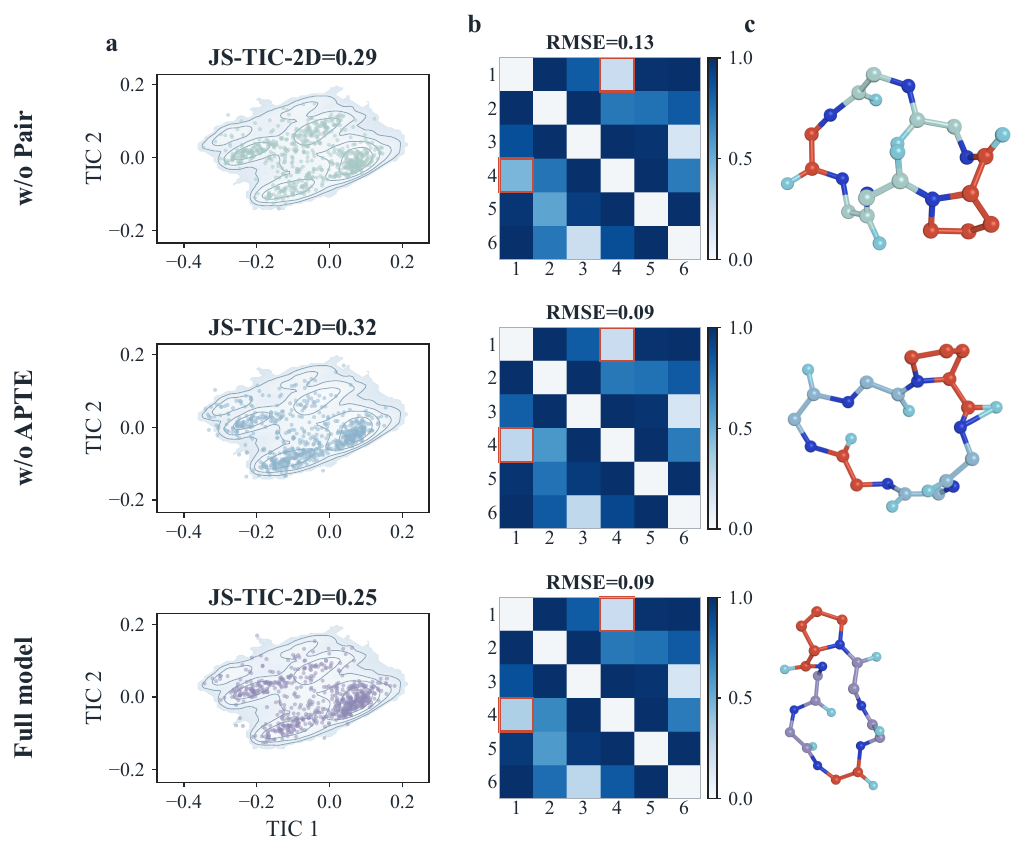}
    \caption{Component-wise case study on GYGPLP.
(a) Generated conformations on the two slowest TICA coordinates; contours show the reference density.
(b) Residue-contact maps, with generated and MD in the lower and upper triangle; red boxes mark the selected pair.
(c) Representative conformations, with those residues highlighted in red.}
    \label{fig:GYGPLP_pair_case}
\end{figure}

\paragraph{Component ablation.}
We ablate the two components behind our contributions, the pair-aware velocity network and APTE, keeping multi-stride sampling in the w/o APTE variant. Removing either degrades every metric, and the two act on complementary aspects of the trajectory (Table~\ref{tab:cyclic_ablation}). Removing the pair representation costs most in spatial geometry, raising JS-PwD from $0.201$ to $0.226$ and JS-Rg from $0.204$ to $0.285$. Removing APTE instead costs most in the slow dynamics, raising JS-TIC from $0.232$ to $0.280$. Both are needed for structural validity, with Val-CA falling from $0.714$ to $0.582$ and $0.544$ respectively, and removing both is worst on every metric.
The same division appears within a single system (Figure~\ref{fig:GYGPLP_pair_case}). For GYGPLP, dropping the pair representation raises the contact-map RMSE from $0.09$ to $0.13$, whereas dropping APTE leaves the contact error unchanged but raises JS-TIC-2D from $0.25$ to $0.32$. APTE therefore governs which regions of the slow landscape are visited, and the pair representation of which residues are in contact.

\begin{table}[t]
    \centering
    \small
    \setlength{\tabcolsep}{5pt}

    \begin{tabular}{lcccc}
        \toprule
        Metric & \methodname{} & w/o Pair & w/o APTE
        & \shortstack{w/o Pair\\and APTE} \\
        \midrule
        \textsc{JS-PwD} $\downarrow$
            & \textbf{0.201} & 0.226 & 0.231 & 0.274 \\
        \textsc{JS-Rg} $\downarrow$
            & \textbf{0.204} & 0.285 & 0.276 & 0.402 \\
        \textsc{JS-TIC} $\downarrow$
            & \textbf{0.232} & 0.252 & 0.280 & 0.287 \\
        \textsc{JS-TIC-2D} $\downarrow$
            & \textbf{0.531} & 0.549 & 0.562 & 0.578 \\
        \textsc{JS-Backbone} $\downarrow$
            & \textbf{0.225} & 0.230 & 0.247 & 0.264 \\
        JS-$\omega$ $\downarrow$
            & \textbf{0.208} & 0.209 & 0.222 & 0.238 \\
        \textsc{Val-CA} $\uparrow$
            & \textbf{0.714} & 0.582 & 0.544 & 0.447 \\
        \bottomrule
    \end{tabular}%
    
    \caption{Component ablation on the cyclic-peptide benchmark.
    Best results are shown in bold.}
    \label{tab:cyclic_ablation}
\end{table}

\paragraph{Multi-timescale trajectory generation.}
To test whether one checkpoint remains reliable across requested temporal resolutions, we evaluate the same model at $s\in\{1,2,4,8,16,32\}$ without stride-specific fine-tuning. 
Within the training range ($s\in[1,16]$), performance is stable across all intervals: the variations in \textsc{JS-TIC-2D} and \textsc{JS-$\omega$} are only (0.0057) and (0.0019), respectively, while \textsc{Val-CA} remains between (0.7027) and (0.7143). 
At the extrapolated stride (s=32), all JS metrics differ from their (s=16) values by at most (0.0013), and \textsc{Val-CA} decreases only slightly from (0.7027) to (0.6963). 
These results show that interval control extends beyond the stride support used for training, rather than only selecting among seen resolutions.
Together, these results directly support the use of the requested interval as an inference-time control for a single trajectory-window generator, which establishes cross-stride robustness of the generated conformational distributions.
Full results, together with the matched-horizon evaluation of kinetic fidelity across sampling intervals, are provided in Appendix C.


\paragraph{Training-interval coverage.}
We next examine whether trajectory generation benefits from broad coverage of temporal scales during training, beyond exposure to only a few discrete intervals.  We train variants on $N\in\{1,2,4,8\}$ fixed strides, leaving all other settings unchanged; the stride sets are listed in Table~\ref{tab:interval_coverage_ablation}.
The full interval distribution is best on five of seven metrics.
In particular, restricting training to a single stride increases \textsc{JS-Rg} from $0.204$ to $0.251$ and reduces \textsc{Val-CA} from $0.714$ to $0.549$. 
Expanding the discrete support from $N=1$ to $N=8$ progressively improves \textsc{JS-PwD}, \textsc{JS-Rg}, and \textsc{JS-TIC-2D}, but none of the sparse schedules matches the full model across the complete metric set except for a $0.001$ advantage on \textsc{JS-Backbone} at $N=8$ and on \textsc{JS-}$\omega$ at $N=4$.
Their lower validity and weaker agreement on the remaining geometric and torsional metrics show that a few isolated interval values are insufficient to recover the full model's overall behavior. 
These results indicate that APTE benefits from a richly sampled temporal axis rather than merely treating one or two interval values as categorical training labels.
We further vary two implementation choices within multi-interval training: the sampling distribution of training strides and the basis used to encode simulation time. Both of them leave the distributional scores essentially unchanged, indicating that the benefit comes from conditioning on the physical timescale itself rather than a particular functional form of the encoding (Appendix C.3).

\begin{table}[t]
    \centering
    \setlength{\tabcolsep}{3.0pt}
    \resizebox{\columnwidth}{!}{%
    \begin{tabular}{lccccccc}
        \toprule
        Training strides
        & \multicolumn{6}{c}{JS distance $\downarrow$}
        & \textsc{Val-CA} $\uparrow$ \\
        \cmidrule(lr){2-7}
        & \textsc{PwD}
        & \textsc{Rg}
        & \textsc{TIC}
        & \textsc{TIC-2D}
        & \textsc{Backbone}
        & $\omega$
        & \\
        \midrule
        $\{2\}$
            & 0.212 & 0.251 & 0.250 & 0.549
            & 0.245 & 0.221 & 0.549 \\
        $\{2,16\}$
            & 0.212 & 0.243 & 0.236 & 0.541
            & 0.228 & 0.210 & 0.603 \\
        $\{1,2,8,16\}$
            & 0.211 & 0.232 & 0.239 & 0.541
            & 0.227 & \textbf{0.207} & 0.643 \\
        $\{1\text{--}6,8,16\}$
            & 0.210 & 0.228 & 0.242 & 0.535
            & \textbf{0.224} & 0.209 & 0.625 \\
        $1$--$16$
            & \textbf{0.201}
            & \textbf{0.204}
            & \textbf{0.232}
            & \textbf{0.531}
            & 0.225
            & 0.208
            & \textbf{0.714} \\
        \bottomrule
    \end{tabular}%
    }
    \caption{Ablation of training-interval coverage on the cyclic-peptide benchmark.
    The full model samples every integer stride in $[1,16]$ approximately uniformly in log space. 
    }
    \label{tab:interval_coverage_ablation}
\end{table}

\subsection{Analysis}
\label{sec:exp_analysis}

\begin{figure}[t]
    \centering
    \includegraphics[width=\columnwidth]{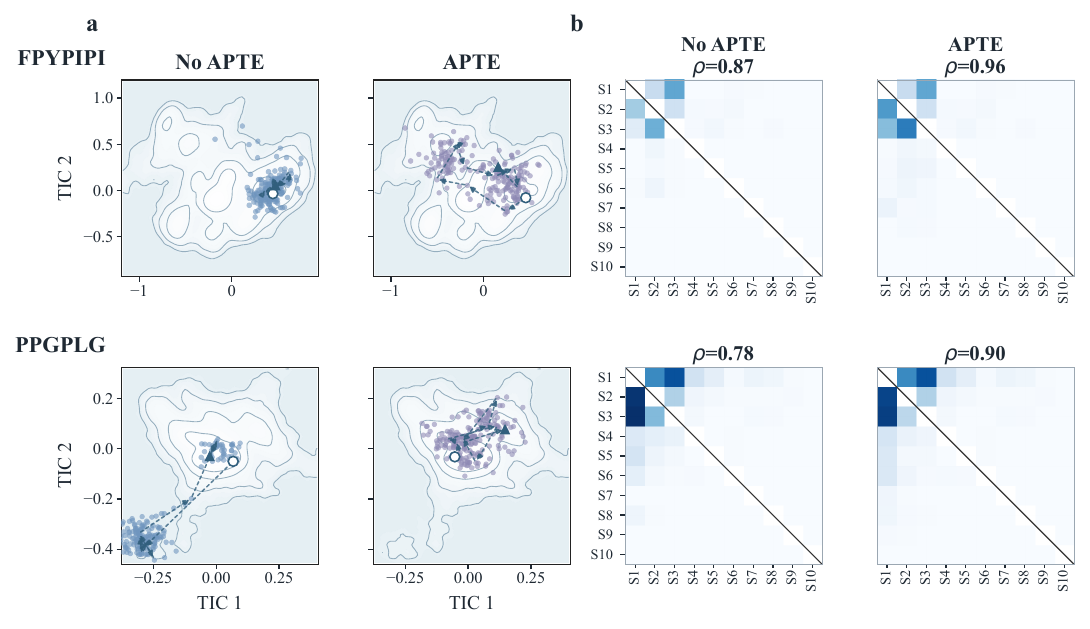}
    \caption{Metastable-state exploration and transition flux for FPYPIPI (top) and PPGPLG (bottom). (a) The first 200 consecutive frames generated without and with APTE, projected onto the two leading reference TICA coordinates; contours show the reference MD density. (b) Symmetrized equilibrium fluxes under reference-defined MSM/PCCA(10) states.
    }
    \label{fig:rpte_metastable_cases}
\end{figure}

\paragraph{Metastable-state coverage and flux.}
Figure~\ref{fig:rpte_metastable_cases} shows the first 200 frames for two cyclic peptides in generated temporal order. Without APTE, FPYPIPI stays near the dominant basin throughout, and PPGPLG makes a sparsely supported excursion into a peripheral region. With APTE, the same number of frames covers a broader portion of the reference-supported landscape and includes intermediates connecting distinct high-density regions. The gain is therefore not merely greater unordered diversity, since transition-relevant conformations appear within a short, temporally coherent segment.
Panel (b) shows the same effect in the equilibrium fluxes, whose off-diagonal rank correlation with the reference rises from $0.87$ to $0.96$ for FPYPIPI and from $0.78$ to $0.90$ for PPGPLG. Both variants are scored under the reference-defined MSM/PCCA decomposition, so these gains reflect a redistribution of flux among identical states rather than a change in the state definition. Full-trajectory projections and additional cases are provided in Appendix~G.1.


\paragraph{Inference efficiency.}
On a single NVIDIA H100, \methodname{} requires $5.33$\,ms per generated
cyclic-peptide conformation, averaged over 1,000-frame trajectories. This
corresponds to a $24.89\times$ speedup over UniSim and a $15.26\times$
throughput improvement over reference MD. Although MDGen and MARS-FM are
slightly faster, \methodname{} achieves substantially stronger generation quality,
yielding a favorable quality--runtime trade-off.
The complete runtime comparison and measurement protocol are reported in Appendix~G.2.


\section{Conclusion}

In this work, we introduced \methodname{}, a multi-scale temporal framework for peptide trajectory generation.
Training windows are cut at strides drawn from a continuous range, and an adaptive physical-time embedding makes the shared velocity field aware of the interval each window spans, while pair-aware geometric features adapt the flow to the flexibility and structural constraints of linear and cyclic peptides.
Across PepMD and a cyclic-peptide benchmark, \methodname{} improves nearly all evaluated metrics and achieves the strongest overall performance on cyclic peptides, with consistent gains in distribution agreement, torsional fidelity, and structural validity at a lower inference cost.
Ablations support the complementary roles of the two designs, and the generated trajectories recover metastable transitions in temporal order.
Modeling multiple temporal scales is therefore an effective design principle for generative models of peptide dynamics.

\bibliography{aaai2027}

\end{document}